# Cosmology and semi-conservation of computations in the universe

Vadim Astakhov
University of California San Diego
astakhov@ncmir.ucsd.edu , vadim_astakhov@hotmail.com

Abstract
Resent works of Hawking and Susskind suggested that information is conserved in the universe. We extend this thesis and propose that dynamics of information - *computations* can conserve in Anti-de-Sitter cosmological model. Information geometry formalism is proposed to analyze information in dynamical systems. We consider entropy flow as a geometrical flow on statistical manifold and develop a *Dynamic Cores* model to analyze migration of information in dynamical systems. Geometrical flow on the statistical manifold was considered as a transition of local dynamical systems in original d+1-dim AdS space to their delocalized holographic representation in d-dim Conformal Field Theory (CFT). It was noted that geometrical flow related to renormalization group flow and provide semi-conservation of informational invariants. Those invariants interpreted as types of computations.

## INTRODUCTION

Term *computation* is a general term for any type of information processing which is changing of information in any manner detectable by an observer. A computation can be seen as a purely physical phenomenon occurring inside a physical system. As such, it is a process which describes everything which changes in the universe, from a particle movement or simple calculations to human thinking.
Let's compare two computer programs running on same computer. One does addition of two numbers 5 and 3 and subtract 2 then return result and another program which add 4 and 2 and subtract 1 then return result. These programs represent two different physical processes with equal amount of information. Moreover the Kolmogorov complexities [1] for processes are equal but the causal structure of information elements is something that makes processes different.
Consider famous example with the astronaut falling into the black hole. Hawking solution [2] suggest that information conserve. But will it guarantee that the astronaut will be able at least to calculate Fibonacci numbers. Information conservation does not immediately guarantee the conservation of informational structures and causal relations among system elements. To resolve this question, we employed holographic principle.
The holographic principle is a speculative conjecture about quantum gravity theories, proposed by Gerard 't Hooft and improved and promoted by Leonard Susskind claiming that all of the information contained in a volume of space can be represented by a theory that lives in the boundary of that region. In other words, if you have a room, you can model all of the events within that room by creating a theory that only takes into account what happens in the walls of the room. One important point is that all local object of original space will have delocalized holographic representation.
Witten result [3] demonstrates that physical system in d+1 AdS can be equivalently described using holographic representation in d-dim Conformal Field Theory. Thus, AdS-CFT is a conjectured duality between string theory in anti-de Sitter space and a conformal field theory on the boundary of anti-de Sitter space.
Based on the fact the conformal field theory is unitary and Wheeler-DeWitt wave function of the universe is time-independent we suggest a hypothesis that string theory might be computation preserving.
We constructed statistical manifold [4] for dynamical systems in AdT/CFT models and demonstrated that principle of information conservation proposed in works of Stephen Hawking and Leonard Susskind can be extended to semi-conservation of computations.
We take entropy flow as a geometrical flow related to renormalization-group flow. Renormalization group is not a formal group but semi-group. Based on Perelman's work [5] we concluded that entropy flow preserve semi-conservation of informational invariants. Such semi-conservation can be interpreted in a sense that universe is open for emergence of new computations but if a computation - "program" once emerge it will never be deleted. This might explain increasing over time complexity of the visible universe.

Delocalization as holography

We propose a concept - "*delocalization*" for computation conservation which is not a direct copying of causal relations but rather their reestablishing due to adjustment in the physical system. As an illustration, consider an example from network theory: where node A sends a message to the node C through the shortest path on the network. That case can be found in many biological systems. The "smart" node A should calculate the shortest paths among network nodes and send message to a proper neighbor. An arbitrary node should be able to perform a few basic operations which we schematically defined as: *get message*, *find next in the path*, and *send message to the next*. As can be easily seen, the system is not functioning if ability to find the shortest path destroyed. But likely, similar functionality can be recovered effectively if each node can be adjusted to perform just simple operations: *get message* and *send message to all neighbors*. The signal will be transmitted from A to C through many paths as well as through the shortest path. We can say that *find the shortest path* function is implicitly emerged in this model as opposed to explicitly implemented in the smart node model.

*Figure1. The left figure represents central-server architecture of communication nodes where a central general server A can sends a request to pass a message to node C through the pre-computed shortest path from A to C. The same functionality can be implemented (right) by broadcasting the message to all neighbors.*

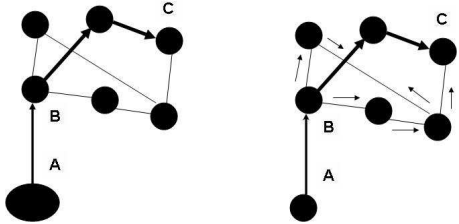

For example, the dynamic that will not broadcast the message but sent it to random node will not lead to emergence of the *shortest path* functionality. This is obviously very noisy example but it illustrates possibility to recover the lost functionality by modification in the system architecture.

The whole idea of *delocalization* seems analogous for well known effect of holographic representation which is well known from various parts of physics. Aside of optical applications there is a mathematical formalism that provide a basis for multiple representations of physical systems. A physical system can be described by holographic representation. That hologram is spatially distributed and requires none-local descriptors. Famous example from mathematical physics is duality between localized in the space physical systems in Anti-de Sitter (AdS) model and their holographic representation in Conformal Field Theory (CFT) [3]. Such holographic- delocalized representation preserve all causal relations of the localized system. It was proposed by Maldacena and demonstrated by Witten that certain conformal field theories in d dimensions can be described on the product of d+1 dimensional AdS space with a compact manifold.
We employ that approach for problem of re-implementation of lost functionality. First, we developed formalism to represent system functionality in geometrical terms. Geometrical formalism naturally merges with AdS formalism and CFT Yang-Mills theory used in work of Witten and others. Then, we suggest localized implementation of the functionality in d+1-dimmensional AdS model and delocalized holographic representation in d-dimensional CFT model. Finally, we consider a process of geometrical flow as a process of delocalization for pre-defined functionality. The process can be interpreted as a process of migration for system functionality from localized to delocalized representation.
Methods from the theory of *dynamical systems* are employed to provide geometrical formalism for analysis of network systems. A *dynamical system* is a mathematics concept in which a fixed rule describes the time dependence of a point in a geometric space. A *system state* is determined by a collection of numbers that can be measured. Small changes in the state of the system correspond to small changes in the numbers. We use the concept of *informational manifold* from "information and statistical geometry" [4] as the geometric *space* and introduce a *system state* as a point on the manifold to describe the dynamics of processes. The numbers are also the coordinates of a geometric space—a manifold. The *evolution* of the dynamical system is a rule that describes what and how future states follow from the current state.

Information Geometry for Analysis of Dynamic Bio-System

We consider a dynamic physical system X composed of n units $\{x_i\}$. Each unit can represent an object like a single neuron or a sub-net of the brain network. Those units can be either "on" or "off" with some probability. "On" means an element contributes to the activity that lead to emergence of the computation Q and "off" is otherwise. Thus observable state of the function $Q = (Q_1, Q_2,..)$ for the system X can be characterized by certain sets of statistical parameters $(x_1, x_2, …)$ with given probability distribution $p(X|Q)$. The idea of endowing the space of such parameters with metric and geometrical structure leads to proposal of use Fisher information as a metric of geometric space for $p(X|Q)$-distributions:
$$g_{\mu\nu} = \int ( \partial p(X|Q) / \partial Q\mu) (\partial p(x|Q) / \partial Q\nu)) p(X|Q) d\{x_i\}.$$
Introduced Fisher metric is a Riemannian metric. Thus we can define distance among states as well as other invariant functional such as affine connection $\Gamma^{\sigma}_{\lambda\nu}$, curvature tensor $R^{\lambda}_{\mu\nu k}$, Ricci tensor $R^{\mu k}$ and Curvature scalar R [4] which describes the information manifold.

The importance of studying statistical structures as geometrical structures lies in the fact that geometric structures are invariant under coordinate transforms. These transforms can be interpreted as modifications of $\{x_i\}$ set by artificial tissue with different characteristics. Thus the problem of a system survival under transition from one bio-physical medium to another can be formulated geometrically.

*Evolution* of the network system can be modeled by Euler-Lagrange equations taken from small virtual fluctuation of metric for scalar invariants. One example of such evolution equation is
$$J = -1/16\pi \int \sqrt{det\ g^{\mu k}}\ (Q)\ R(Q)\ dQ.$$
But for open system such as neural network, external constraint can be added as a scalar term dependent on arbitrary covariant tensor $T^{\mu k}$:
$$J = -1/16\pi \int \sqrt{g(Q)}\ R(Q)\ dQ + 1/2 \int \sqrt{g(Q)}\ T^{\mu k}\ g_{\mu k}\ dQ.$$
That lead to well known geometrical equation $R^{\mu k}(Q) - g^{\mu k}(Q)R(Q) + 8\pi T^{\mu k}(Q) = 0$ which describe metric evolution. Solutions of this equation represent statistical systems under certain constraints defined by tensor $T^{\mu k}$. Thus, functionality of network systems can be resented in terms of gravitation theory.

Another way to employ geometrical approach is to define tangent vector "A" for each point X of manifold M(X) as: $A_\mu \sim \partial \ln(p(x))/\partial x_\mu$
Where Li brackets $[A_\mu A_\nu] \sim A_k$, give use the way to find transformations which will provide invariant descriptors. First we can employ approach developed in gauge theories that are usually discussed in the language of differential geometry that make it plausible to apply for informational geometry. Mathematically, a *gauge* is just a choice of a (local) section of some principle bundle. A gauge transformation is just a transformation between two such sections. Note that although gauge theory is mainly studied by physics, the idea of a connection is not essential or central to gauge theory in general. We can define a gauge connection on the principal bundle. If we choose a local basis of sections then we can represent covariant derivative by the connection form $A_\mu$, a Lie-algebra valued 1-form which is called the gauge potential in physics. From this connection form we can construct the curvature form *F*, a Lia-algebra valued 2-form which is an intrinsic quantity, by $F_{\mu\nu} = \partial_\mu A_\nu - \partial_\nu A_\mu - ig[A_\mu\ A_\nu]$
$[D_\mu D_\nu] = -ig F_{\mu\nu}$, where $D_\mu = \partial_\mu - igA_\mu$; $A_\mu = A_\mu^a t^a$ and t –is generator of infinitesimal transformation
Thus we can write Lagrangian: $1/4\ F_{\mu\nu}^a F_{\mu\nu}^a \Leftrightarrow -1/2\ Sp(F_{\mu\nu} F_{\mu\nu})$ that is invariant under transformation of coordinates. Such approach provides us with informational analog of CFT Yang-Mills model.

At the same time, another evolution functional $J= \int (R+|\nabla f|^2)exp(-f)dV = \int (R+|\nabla f|^2)dm$, that is dependant on function f –gradient vector field defined on the manifold of volume V. It is well known in super-gravity (and string theory) [5] functional that can lead us to AdS model. Thus using statistical geometry approach, we can formulate network system functionality in terms of two models: super-gravity and Yang-Mills.

It can be easely shown that functional $J=\int (R+|\nabla f|^2)dm$ can be taken as the gradient flow $dg_{ij}/dt =2(R_{ij}+ \nabla_i \nabla_j f)$ that is generalization of the geometrical flow or so called Ricci flow $dg_{ij}/dt = -2R_{ij}$. Interesting thing about Ricci flow is that it can be characterized among all other evolution equations by infinitesimal behavior of the fundamental solutions of the conjugate heat equation. It is also related to the description of the renormalization group flow.

## Dynamic Core of a function

We introduce a definition *Dynamic Core* -- A dynamic system that consists of a set of dynamic elements causally interacting with each other and the environment in a way that lead to a high level of information integration within the system and emergence hierarchical causal interactions within the system. A higher level of causal power among system elements is compared to causal interactions with an environment. The dynamic core can be seen as a functional cluster characterized by strong mutual interaction among a set of sub-groups over a period of time. It is essential that this functional cluster be highly differentiated.

Dynamic Core (DC) will be used to describe any dynamical system that has a sub-part acting in causal relations with each other. To measure *causal relation* some metrics considered are *information integration* and *causal power*. Dynamic core is defined as a sub-system of any physical environment that has internal *information integration* and *causal power* [6-7] much higher than mutual *information integration* between the system and the environment.

To describe specialized sub-networks relevant to emergence of specific high level functions, we employ concept of functional cluster [9]: If there are any causal interactions within the system such as signals transfer, the number of states that the system can take will be less than the number of states that its separate elements can take. Some sub-nets can strongly interact within itself and much less with other regions of the brain. Geometrically, it is equivalent to higher positive curvature $R \sim CI(X^k_j) = I(X^k_j) / MI(X^k_j; X - X^k_j)$ [9] in the area of information manifold which reflects the system $X^k_j$ state dynamics due to the loss of information entropy "H". The loss is due to interactions among the system elements - $I(X^k_j) = \sum H(x_i) - H(X^k_j)$ and interaction with the rest of system described by mutual entropy $MI(X^k_j; X - X^k_j)$.

*Figure 2. System X partitioned to subset of elements $X^k_j$ and the rest of the system $X - X^k_j$. The dashed ellipse represents another possible partition. The dependence between the subset $X^k_j$ of k elements and the rest of the system $X - X^k_j$ can be expressed in terms of mutual information: $MI(X^k_j, X - X^k_j) = H(X^k_j) + H(X - X^k_j) - H(X)$*

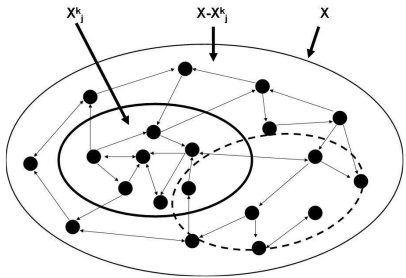

Curvature R near 1 indicates a subs-net which is as interactive with the rest of the system as they are within their subset. On the other hand, a R much higher than 1 will indicates the presence of a *functional cluster*- a subset of elements which are strongly interactive among themselves but only weakly interactive with the rest of the system. Function cluster is a manifestation of specialized region that involve in generation of high-level function. Geometrically, it will emerge as a horn area on information manifold.

Evolution of causal interactions among functional clusters is described by Ricci tensor $R_{ij}$ which is geometric analog to the concepts of *effective information* and *information integration* [6]. Effective information $EI(X^k_j \rightarrow X - X^k_j)$ between sub-net $X^k_j$ and $X - X^k_j$ can be seen as an amount of informational entropy that $X - X^k_j$ shares with $X^k_j$ due to causal effects of $X^k_j$ on $X - X^k_j$.

## Geometric Flow as Delocalization and Functionality Conservation

Summarize previous chapters, we propose model where an evolution of the system and the *architecture transition can be seen as a geometrical flow of information entropy on some informational manifold*. Results [5] demonstrates that Ricci Flow can be considered as renormalization semi-group that distribute informational curvature over the manifold but keep invariant $R=R_{min}*V^{2/3}$ where R-curvature and V-volume on information manifold. Region with strong curvature interpreted as sub-system with high information integration and recursive complexity. Thus, we proposed Ricci flow as a process of delocalization that provides distributed representation under architecture transition. Based on Perelman works [10] for the solutions to the Ricci flow ($d/dt\, g_{ij}(t) = -2R_{ij}$) the evolution equation for the scale curvature on Riemann manifold:

$$d/dt\, R = \Delta R = 2\,|Ric|^2 = \Delta R + 2/3\, R^2 + 2|Ric^o|^2$$

It implies the estimate $R^t_{min} > -3/(2*(t+1/4))$ where the larger t-scalar parameter then the larger is the distance scale and smaller is the energy scale.

The evolution equation for the volume is $d/dt\, V < R_{min}V$. Take R and V asymptotic at large t, we have $R(t)V(t)^{-2/3} \sim -3/2$. Thus, we have Ricci flow as a process of delocalization where V-growth when R-decrease that provides distributed representation under architecture transition.

*Figure 3. Demonstrate delocalization of functional cluster-dynamic core due to geometry flow*

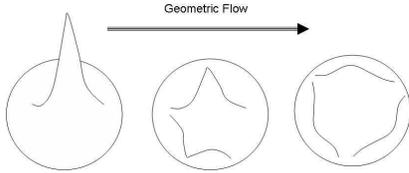

Another evolutional equation that can be introduced as a candidate for delocalization is Calabi flow which, unlike the Ricci flow, is only defined on Kahler manifolds with complex coordinates ($z_i, z'_j$):
$\partial g_{ij}/\partial t = \partial R/\partial z_i \partial z'_j$. A geometrical context for the Calabi flow represents spherical waves of informational metrics.

*Figure 4. Demonstrate distributed representation of a Dynamic Core on 3D informational manifold as a set of concentric conics representing individual points. Each point of the Dynamic Core is a state of an information system and the object itself is an assembly of causally related information objects.*

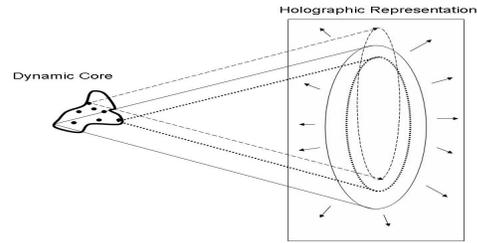

Define informational metric as $g_{ij} = 2(\exp(\Phi(z,z',t)))$ – exponent of a scalar then two flows assume the same form Ricci: $\partial\Phi/\partial t = \Delta\Phi$ and Calabi: $\partial\Phi/\partial t = -\Delta\Delta\Phi$

## Coherent structure Pfaff Dimension and Topological Torsion as a self poetic functionality

We also investigate topological evolution of informational manifold. Dynamic core is viewed as a coherent structure such as deformable connection domain on information manifolds with certain similar topological properties. We consider topological properties of such space that stay the same under continuous transformations. For the scalar function X evolves in time with some velocity VX, following the method of Cartan [11], a certain amount of topological information can be obtained by the construction of the Pfaff sequences based on the 1-form of Action, $A = ds = VX_\mu(x)dx^\mu$ a differential constructed from the unit tangent information integration velocity field VX. We claim that emergent states are coherent topological structures:

*Topological Action (energy) A*
*Topological Vorticity (rotation) F=dA*
*Topological Torsion (entropy) H=A ∧ dA*
*Topological Parity (Dynamic Core) K = dA ∧ dA*

The rank of largest non-zero element of the above sequence gives Pfaff dimension of an information manifold. It gives us the minimal number M of functions required to determine the topological properties of the given form in a pre-geometric variety of dimensions N. We require at least dimension 4 to accommodate complex systems with dynamic cores.

It may be demonstrated from deRham theorem and Brouwer theorem [6] that the odd dimensional set (1,3,..) may undergo topological evolution but even dimensions remain invariant. It implies that Dynamic Core coherent topological structure once established through evolution of the Pfaff dimension from 3-to -4 then will remain invariant. Also, the Pfaff dimension is an invariant of a continuous deformation of the domain thus it is invariant under geometrical flow. We developed one particular example of such transition using Ising model [Appendix A].

**CONCLUSION**

We analyze and prove the thesis that our universe can provide conditions not just for information conservation but also for semi-conservation the information computations.

"Dynamic Core conservation Hypothesis" was proposed that state: Some dynamic systems-universes can exhibit certain level of complexity and symmetry that manifested as a conservation of emergent dynamic structures. Dynamic core once emerged can not be destroyed (Tuff topological invariant). Mathematically such conservation of emerged "DC" is equivalent to the semi-group flow that provides forward in time conservation but does not provide reverse conservation. We can call it as a week conservation low to distinguish from strong conservation lows like conservation energy or information.

Information geometry formalism used to analyze conservation of physical computations in dynamic physical systems. It was suggested, that an arbitrary computations can be represented in form of Conformal Yang-Mills field theory (CFT) as well as in form of supergravity (and string) theory. Those models can be mathematical formulated in terms of informational geometry.

Paper [3] demonstrates precise correspondence between formalisms of d-dimensional CFT and supergravity on the product of d+1-dimensional AdS space with a compact manifold. AdS model is assigned to local d+1-dim implementation of computations by specialized physical system. On the other hand, d-dim CFT is suggested for holographic (delocalized) implementation of the same computations. Geometrical flow on informational manifold is suggested as a mechanism of transition from one representation to another. Such transition can conserves topological properties and preserve causal relations of the computational dynamic core.

In some sense, our formalism can be a physical toy model for Graham Ward 'emergent dualism' or qualified monism whereby the computations generated through a physical processes and is a set of causal relations developed over a live time but it is capable of surviving its original processes dissolution by migrating to another physical host.

Joking back about our astronaut who stuck into the back, it is still unclear can his mind and consciousness survive [5-6] but it seems that computational processes of his brain will continue to run.

## [Appendix A] Matrix theory and Ising mode for Information Geometry

Ricci flow was proposed as an information entropy flow that conserves dynamic core functionality by providing holographic representation for the initially localized DC. Also we proposed a toy model for holographic representation which is based of Smolin [9] matrix model with hidden variables.

Any holographic representation sub-system $X_a^b$ can be seen as a matrix $N*N$ –of N of informational objects, where diagonal "ii" represent holographic set of projected eclipses and internal information integration for "i" if it has internal structure. And off-diagonal element "ij" represent effective information between elements i and j.

Interactions among sub-systems can be expressed by an Action $S \sim m \int dt\, Tr\{X'^2_a + \omega^2[X_a X_b][X^a X^b]\}$ where $X_a$ is $N*N$ matrix that can be represented $X=D+Q$ as sum of diagonal $D=diag(d_1,d_2,..)$ and none-diagonal pieces. Then action can we wrote as $S \sim m \int dt\, (L^D + L^Q + L^{int})$

Nelson's stochastic formulation of quantum theory emerges naturally as a description of statistical behavior of the eigenvalues with interaction potential of interaction between diagonal and none-diagonal elements:

$$L^{int} = U(D,Q)$$
$$L^D = m \sum D'^2_a$$
$$L^Q = m\{\sum Q'^2_a + \omega^2 [Q_a Q_b][Q^a Q^b]\}$$

$L^{int} = 2m\omega^2 \sum \{-(d_i-d_j)^{2a} Q^2_b - (d_i-d_j)_a (d_i-d_j)_b Q^a Q^b - 2(d_i-d_j)\,^a Q^b [Q_a Q_b]\}$ based on this potential the classical equation of motion can be wrote how each matrix element moves in an effective potential created by the average motion of other elements. We assume that statistical averages satisfy (Gaussians processes) relations consistent with the symmetry of the theory. That gives use Brownian movements in potential:

$$<U> = m\Omega_Q^2/2\, Q_{ija}\, Q^{ija} + m\Omega_d^2/2\, (d_a^i - d_a^j)^2$$
$$\text{where } \Omega_Q^2 = 4(d-1)\omega^2 [(N-1)q^2 + 2r^2]$$
$$\Omega_d^2 = 4(d-1)\omega^2 q^2$$

The Q system is in distribution. Variation principle for Matrix Model can be reformulated for eigenvalues. $\lambda = d + \sum Q^{ij}_a Q^{ji}_a/(d_i-d_j)_a + \ldots$. And Diffusion constant for the eigenvalues is $\nu = (\Delta d)^2/\Delta t$

Now we would like to show that neural-network can emulate quantum mechanical system at normal temperature. We are going to make one assumption that $T/(8(d-1)m\omega^2) = t/N^p$ and $\hbar = m\nu$, then we can define wave function: $\Psi = \sqrt{\rho} \exp(S/\hbar)$, where $\rho$ - probability density $= 1/Z \exp(-H(Q)/T)$ and Hamiltonian $H(Q) = m\{\sum Q'^2_a - \omega^2[Q_a Q_b][Q^a Q^b]\}$.

That illustrate Lee Smolin result that variation principle in presence of Brownian motion equivalent to Schrödinger equation:

$$i\hbar\, d\Psi/dt = \{-\hbar/2m\, d^2/d(\lambda)^2 + m\Omega_d^2/2 \sum (\lambda_a^i - \lambda_a^j)^2 + TN(N-1)/4 + NmC\}\,\Psi$$

The analysis provides analogy between neural network dynamics and effects of quantum theory. Now, we have Hamiltonian to try Ising approach for the information geometry approach. As we demonstrated in the paper, the geometric structure in which informational manifold is endowed leads to certain local invariants, one of the most important being the Ricci scalar curvature R. $R \sim \zeta^d$, where $\zeta$-is the correlation length, which is two-point function, and d- denotes the number of spatial dimensions of the model. The curvature also receives contribution from higher order correlations.

The scaling behavior of the curvature in the vicinity of the critical point provides a satisfying picture of how certain universal features of the near-critical regime are encoded in the Fisher-Rao geometry of the informational manifold.

We will call $N^*$-critical size, above which the system is to a reasonable approximation 'thermodynamic'. When the size drop below $N^*$ the system often behaves in qualitatively different way. R is strictly positive in the thermodynamic regime and negative at none-thermodynamic regime.

We can start with the state of the system immersed in a large heat bath with fixed temperature T in thermal equilibrium. The Gibbs measure can be mapped to Hilbert space $P(x|\theta) = \exp(-\sum_{i=1:r} \theta_i H^i(x) - \ln Z(\theta)) = \psi_\theta(x)$ where $\{\theta_i\}$-represent r-dim sub-space S in the Hilbert space. The Fisher–Rao metric can be introduced on the maximum entropy manifold as : $g_{ij} = 4\int \partial_i \psi_\theta(x)\, \partial_j \psi_\theta(x)\, dx = \partial_i \partial_j \ln Z(\theta) = \partial_i \partial_j S(\text{entropy})$, $\partial_i = \partial/\partial \theta^i$.

And an entropy $S(p|q) = \int p\, \ln(p/q) = S(p(\theta)|p(\theta+d\theta)) = 1/2\, g_{ij}\, d\theta^i\, d\theta^j + \ldots$ is represented as Taylor series. Then manifold can be interpreted as a maximum entropy surface and consequently specific geodesics will correspond to equations of state for the system.

We take simple (Ising chain) source-to-source (H1) and source to net (H2) interaction then r=2 and $-\beta H = \beta \sum s_i s_j + h \sum s_i$, $\beta = 1/kT$, $s=\{+1,-1\}$, h-network "field" then Scalar curvature

$R = -1/(2\det(g)) * \det\{$
$\partial_1^2 \ln Z \quad \partial_1 \partial_2 \ln Z \quad \partial_2^2 \ln Z$

$$\begin{matrix} \partial_1^3 \ln Z & \partial_1^2 \partial_2 \ln Z & \partial_1 \partial_1^2 \ln Z \\ \partial_1^2 \partial_2 \ln Z & \partial_1 \partial_1^2 \ln Z & \partial_2^3 \ln Z \end{matrix}$$
}
will act as an indicator of finite size effects. Differentiate $N^{-1}$ in Z we have
$g_{ij} = 1/N \, \partial_i \, \partial_j \, \{N \beta + \ln [(\cosh h + \eta)^N + (\cosh h - \eta)^N]\}$, where $\eta = (\sinh^2 h + \exp(-4\beta))^{1/2}$

Thus the size of dynamic core depends on temperature as T~size.
Through Ricci flow R<0 (localized DC) can transit to R>0 delocalized implementation.
 if N->∞ then we have thermodynamic curve $R = 1 + \eta^{-1} \cosh h > 0$.

Figure 5. Process of Holographic representation can be seen as Geometric flow that smooth and delocalize curvature. Region with strong curvature interpreted as sub-system with high information integration and recursive complexity

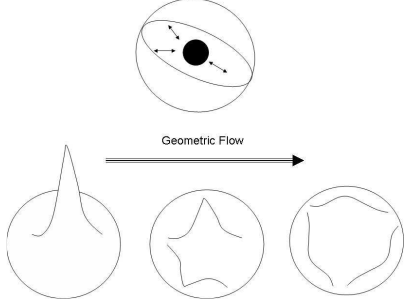

Now, lets consider just two sources with interaction described by a Hamiltonian: $H = \sum C_i * \sigma_i * \tau_i$.
The prescribed state is denoted by the density matrix $\rho = |\psi\rangle\langle\psi|$, where $|\psi\rangle = \cos(\theta/2)|0\rangle + \exp(i\varphi)\sin(\theta/2)|1\rangle$ is superposition of active $|1\rangle$ and $|0\rangle$ none-active state. As we can see an initial state of the system of two sources is $|\rho(0)|_{4\times 4}$. The information encoded in the system at time t is characterized by the fidelity $F_i(t) = \langle\psi | \rho^i(t) | \psi\rangle$, where $\rho^i(t) = \text{Tr}\, \rho(t)$-except "i".
Let's introduce two qualities:
$$CF_i(t) = \cos^2(\theta/2) \rho^i_{00}(t) + \sin^2(\theta/2) \rho^i_{11}(t)$$
$$QF_i(t) = \text{Re}[\exp(-i\varphi) * \sin(\theta) \rho^i_{10}(t)]$$

And rewrite Fidelity as $F_i(t) = CF_i(t) + QF_i(t)$, where $\rho(t) = \exp(-iHt) * \rho(0) * \exp(iHt)$.
By straightforward calculations [14], we found that if $C_i = C_j$ (uniformity of information integration among all sources) for any i and j then $\sum CF_i(t)$ and $\sum QF_i(t)$ are both invariable. That mean total $F(t) = \sum F_i(t)$ is invariable too.
Due to the interactions between parts of the system, the states of these parts changes with time, and the information is expanded between them. But the total Fidelity-information is conserved. This is something like Energy Conservation Law for information systems. This law leads to interesting phenomena when information can partially concentrated spontaneously in one part of the whole system due to oscillation part of $F_i(t)$. Such concentration and following dilution in large-scale system with thermal bath can be seen as localization and de-localization of the Dynamic Core.
The theorem [10] also gives us a mechanism to construct informational systems (universes) with conservative dynamic core. Such universes should have "conservative Hamiltonian" that will preserve total information integration. Hamiltonian should satisfy condition $[H, \sum_{i=1}^n C_i] = 0$, where $C_i = 1/(2^{n-1}) I_1 * I_2 * \ldots \rho^0_i * I_{i+1} * I_{i+2} * \ldots I_n$, with $\rho^0_i$ denotes the reduced density matrix of the i-th source and $I_i$ denotes reduced density matrix of other sources. This is easily can be proven by showing that $\partial F(t)/\partial t = 0$ is equal to $[H, \sum_{i=1}^n C_i] = 0$.